\documentclass[amsmath,amssymb,aps,prl,twocolumn,floats,superscriptaddress]{revtex4-2}

\usepackage{graphicx}
\usepackage{amssymb}
\usepackage{amsmath}
\usepackage{caption}
\usepackage{placeins}
\usepackage{hyperref}
\usepackage{footnote}
\usepackage{color}
\usepackage{multirow}
\usepackage{enumerate}
\usepackage{soul}
\usepackage{xcolor}
\hypersetup{
  colorlinks   = true, 
  urlcolor     = magenta, 
  linkcolor    = magenta, 
  citecolor   = magenta 
}

\pdfoutput=1 
\begin{document}
 
\title{Chiral phonons driven by chiral cavities} 


\author{V. A.S.V. Bittencourt}
\email{bittencourt@fisica.unam.mx}
\affiliation{Instituto de F\'{i}sica, Universidad Nacional Aut\'{o}noma de M\'{e}xico, Mexico City, Mexico.}
\affiliation{Institut de Science et d’Ingénierie Supramoléculaires (ISIS, UMR7006), Universit\'{e} de Strasbourg, 67000 Strasbourg, France}
\author{N. Shabala}
\affiliation{Department of Physics and Astronomy, Chalmers University of Technology, 412 96 Göteborg, Sweden}  
\author{R. M. Geilhufe} 
\affiliation{Department of Physics and Astronomy, Chalmers University of Technology, 412 96 Göteborg, Sweden}
\author{A. Metelmann} 
\email{anja.metelmann@kit.edu}
\affiliation{Institut de Science et d’Ingénierie Supramoléculaires (ISIS, UMR7006), Universit\'{e} de Strasbourg, 67000 Strasbourg, France}
\affiliation{Institute for Theory of Condensed Matter and Institute for Quantum Materials and Technology, Karlsruhe Institute of Technology, 76131, Karlsruhe, Germany}
\date{\today}
\begin{abstract}
Lattice vibrations carrying angular momentum give rise to fundamental phenomena such as the phonon Hall effect and the Einstein–de Haas effect, offering new opportunities for manipulating angular momentum in condensed-matter systems.
Generating such circularly polarized phonons requires either the use of an external magnetic field, or of polarized light pulses. The latter approach does not require the use of any magnetic response of the material, but it has limitations, in particular regarding the duration of the pulse. Consequently, any effect stemming from phonons generated by light pulses is transient. In this paper, we propose the use of a driven electromagnetic cavity as a route to generate a steady-state population of chiral phonons. We derive a general description of cavity-chiral phonon interaction, with a particular focus on modes of chiral cavities. We show that, under optimized conditions, a significant effective phonon-induced magnetic field can be generated by means of an external drive with realistic power. Our approach is specially tailored for $\Gamma$ point phonons with THz frequencies, and opens a new route for investigating chiral phonons with electromagnetic cavities.
\end{abstract}

\maketitle

\section*{Introduction}

\begin{figure}
    \centering
    \includegraphics[width=0.75\columnwidth]{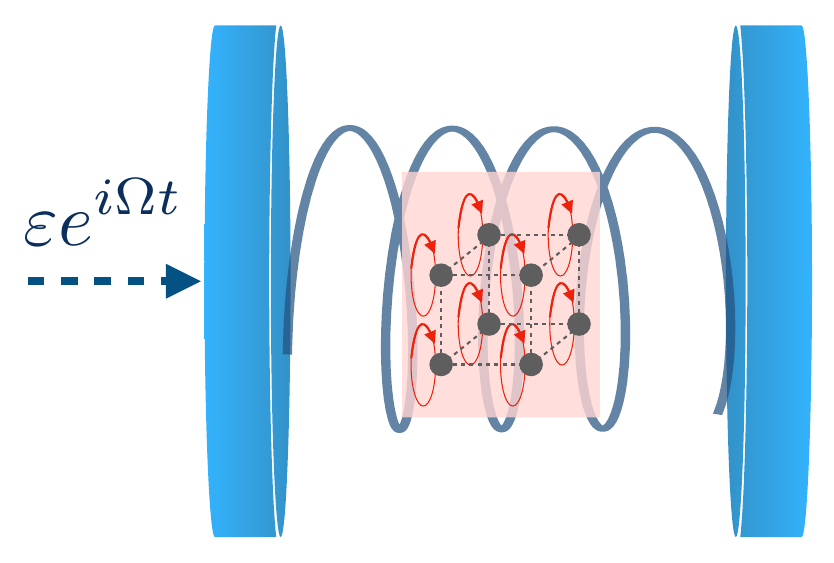}
    \caption{Chiral Phonons generated by a coherently driven chiral cavity (in blue). The material hosting chiral phonons, associated with the circular motion of the crystal's ions (represented here illustratively), is loaded in a chiral cavity, and supports phonon modes with a definite handedness (in red). The cavity mode is coherently driven with a monochromatic tone of amplitude $\varepsilon$ at frequency $\Omega$.}
    \label{fig:Scheme}
\end{figure}
\vspace{0.5cm}

Phonons with a non-zero angular momentum present a pathway to induce magnetization through the phono-magnetic effect \cite{shabala2025axial}. Such phonons are referred to as axial, or circularly-polarized, or, if they break improper rotational symmetry, chiral phonons \cite{juraschek2025chiral}. 
The existence of the phono-magnetic effects has been supported by a wide range of experimental observations~\cite{Schaack1975, Schaack1976, Schaack1977, baydin2022magnetic, hernandez2023observation, lujan2024spin, mustafa2025origin, che2024magnetic, cheng2020large,tang2024exciton, wu2023fluctuation, wu2025magnetic, basini2024, luo2023large,Davies2024} and theoretical predictions~\cite{juraschek2017dynamical,Juraschek2019OrbitalPhonons,Geilhufe2021DynamicallyKTaO_3,ren2021phonon,shabala2024phononinverse,chaudhary2024giant,klebl2024ultrafast,Merlin2024UnravelingFields,Merlin2024MagnetophononicsMisnomer}.

The phono-magnetic effect arises through two mechanisms~\cite{shabala2026unifying}.
One of them is the dynamical multiferroicity, i.e. magnetization induced by the circular motion of ions through the change of polarization~\cite{juraschek2017dynamical,Juraschek2019OrbitalPhonons,Geilhufe2021DynamicallyKTaO_3}.
The second contribution arises from the electron-phonon coupling which causes splitting of the electronic energy levels.
The size of the splitting is proportional to the effective magnetic field, which acts as a time-reversal symmetry breaking field inside of the material~\cite{Merlin2024UnravelingFields, Merlin2024MagnetophononicsMisnomer, shabala2024phononinverse, shabala2026unifying}.
The focus of this paper is the second mechanism, in particular, the effective magnetic field.
Multiple microscopic theories describing the effect highlight the quadratic dependence of the size of the splitting, and hence, the effective magnetic field, on the phonon amplitude~\cite{shabala2026unifying}.
In quantized form, this dependence is manifested as a population imbalance between right (R)-  and left(L)-circularly polarized phonons \cite{shabala2024phononinverse,Merlin2024MagnetophononicsMisnomer}.

An imbalance between polarized phonons can be generated by magnetic means, 
via the coupling to magnetic order \cite{che2024magnetic}, magnetic excitations \cite{an2020coherent,an2022bright,muller2024chiralphonons}, or by applying a circularly polarized laser \cite{luo2023large,Davies2024,basini2024,Biggs2025UltrafastLiNbO3,Minakova2025DirectModes}. 
In the first case, suitable for systems that exhibit a ferro- or paramagnetic responses, the bias field generates a magnetization, whose excitations can then generate chiral phonons \cite{an2020coherent,an2022bright,muller2024chiralphonons}. 
In the absence of an external bias field, circularly polarized light pulses can be used to excite chiral phonons resonantly. In such a scenario, the imbalance between circularly polarized phonons is transient -- after the light pulse is over, the generated population decays. While longer pulses could be used to generate more persistent phonon population, it would also incur undesired side-effects, such as excessive sample heating.

In this paper, we propose an alternative scheme for generating a steady-state imbalance between left and right-circularly polarized phonons with light that does not rely on pulsed excitations. We consider the architecture shown in Fig.~\ref{fig:Scheme}: the material hosting the phonons is loaded in a cavity, which confines light. The cavity is externally driven with a laser, such that the driven-dissipative dynamics of the system yields a steady-state with a finite phonon population. Related cavity-based approaches have been focused on controlling phonon excitations with multicolor pulses \cite{Yaniv_cavity}, while here we focus on a coherently driven chiral cavity to generate an imbalance between circularly polarized phonons. We specifically consider the use of chiral cavities which exhibit eigenmodes with fields of definite (circular) polarization \cite{hubner2021}. These can be engineered with structured mirrors and other metamaterial structures \cite{Voronin2022,forero2024,rebholz2025}, and can be designed to have well isolated modes at target frequencies. We model the cavity-phonon coupling for cavities modes with frequencies matching that of the phonon modes, and obtain a cavity-phonon Hamiltonian that describes the coherent exchange of excitations between photons and phonons. In state-of-the-art cavities with a rectangular geometry, and for samples of $~10 \, \mu {\mbox{m}}^3$, effective static phonon magnetic fields of $~4$ mT can be generated with a drive power $\sim 10$ nW. This is comparable with effective fields generated with moderate THz pulses \cite{luo2023large,Davies2024,basini2024,Biggs2025UltrafastLiNbO3}, albeit being weaker than the values reported in \cite{Biggs2025UltrafastLiNbO3} due to the strong dissipation of the $\Gamma$-point phonons considered here. Our results show that cavity-based systems promissing route for future investigation of axial phonons and their manipulation without requiring the use of external biases field and relying on magnetic properties.

\section{$\Gamma$ point phonons and effective charge: coupling to the electromagnetic field}

In ionic crystals, displacements of the ions from their equilibrium positions changes the electrical polarization of the material due to the ions' charged nature. This in turn modifies how the crystal responds to an external electromagnetic field, yielding the coupling mechanism between light and phonons. Phonons are described by the displacement $\vec{u}_{l \kappa}$ of the ion $\kappa$ hosted in the unit cell $l$ from its equilibrium position $\vec{R}_{l \kappa}^0$:
\begin{equation}
    \vec{u}_{l \kappa} = \vec{R}_{l \kappa} - \vec{R}_{l \kappa}^0.
\end{equation}
To first order in the displacement, the electric polarization density vector of the crystal is given by
\begin{equation}\label{eq:polarization}
    P_{\alpha} = \sum_{l \kappa  \beta} \frac{\partial P_{\alpha}}{\partial u_{l \kappa, \beta}} u_{l\kappa, \beta},
\end{equation}
where $\alpha$ and $\beta$ denote the Cartesian components of each vector. Here, we assume that the electric polarization of the crystal vanishes when atoms are at their equilibrium position. 

To model the coupling between phonons and photons, we use the Born effective charge \cite{gonze1997dynamical, ghosez1998dynamical}
\begin{equation}
\label{eq:effective_charge_born}
    Z_{\kappa, \alpha \beta} = V_0 \frac{\partial P_\alpha}{\partial u_{l \kappa \beta}},
\end{equation}
which associates the polarization induced by the displacement of the ions to an effective electric charge. Here, $V_0$ denotes the unit cell volume.
The displacement $u_{l \kappa \beta}$ can then be written in terms of the creation and annihilation operators, $\hat{a}_{\vec{q}, \nu}$, for phonons with wave-vector $\vec{q}$ and polarization $\nu$ as \cite{giustino2017electron}
\begin{equation}\label{eq:u_quantized}
    u_{l \kappa \beta} = \left(\frac{m_0}{N_l m_\kappa}\right)^{1/2} \sum_{\vec{q}\nu} e^{i\vec{q}\cdot \vec{R}_l} \xi_{\kappa \beta \nu} l_{\vec{q}\nu} (\hat{a}_{\vec{q}\nu}^\dagger+\hat{a}_{\vec{q}\nu}),
\end{equation}
where $m_0$ is an arbitrary reference mass, $N_l$ is the number of unit cells and $\xi_{\kappa \beta \nu}$ denote the phonon polarization vectors. While Eq. \eqref{eq:effective_charge_born} describes the effective charge with respect to an ion and the Cartesian directions, it is also possible to write down the effective charge for the phonon modes: $Z_{\nu \alpha} (\vec{q}) = \sum_{\kappa \beta} Z_{\kappa \alpha \beta} \frac{\xi_{\kappa \beta \nu}(\vec{q})}{\sqrt{m_\kappa}}$. With Eq. \eqref{eq:u_quantized} and the definition of the mode effective charge, we can write down the polarization operator as
\begin{equation}
\label{eq:Pol}
    \hat{P}_{\alpha} = \frac{1}{V_0} \left( \frac{m_0}{N_l}\right)^{1/2} \sum_{\vec{q}\nu} e^{i\vec{q} \cdot \vec{R}_l} Z_{\nu \alpha}(\vec{q})l_{\vec{q}\nu}(\hat{a}_{\vec{q}\nu} + \hat{a}_{\vec{q}\nu}^{\dagger}).
\end{equation}

In what follows, we focus on the coupling between cavity modes and $\Gamma$-point phonons. Those are modes localized at the center of the Brillouin zone, and have vanishing momentum. 
In crystals such as SrTiO$_3$ and KTaO$_3$ such phonon modes have frequencies in the THz range \cite{vogt1995refined}. The polarization of a $\Gamma$-point phonon mode yields
\begin{equation}\label{eq:polarization_quantized}
        \hat{P}_{\alpha} = \frac{1}{V_0} \left( \frac{m_0}{N_p}\right)^{1/2} Z_{\nu \alpha}l_{\vec{0}\nu}(\hat{a}_{\vec{0}\nu} + \hat{a}_{\vec{0}\nu}^{\dagger}),
\end{equation}
with $l_{\vec{0}\nu} = \sqrt{\frac{\hbar}{2\omega_{\vec{0}\nu}m_0}}$. Depending on the crystal structure, more than one mode can be supported at the $\Gamma$-point. 
For example, in SrTiO$_3$ and KTaO$_3$ crystals, there are two orthogonal linearly polarized modes at the $\Gamma$-point. Such modes can then combine, forming circularly polarized phonons that exhibit definite handedness \cite{basini2024, geilhufe2023electron}.

\section{Coupling to a cavity mode}

The coupling between phonons and photons in a cavity can be obtained from the energy density $w_{\rm{EM}}$ of the electromagnetic field in a linear and non-dispersive medium, which is given by \cite{landau}
\begin{equation}
w_{\rm{EM}} = \frac{1}{2} \left(\vec{E}\cdot \vec{D} + \vec{B}\cdot \vec{H} \right).
\end{equation}
The electric displacement vector $\vec{D} = \vec{P} + \epsilon_0 \vec{E}$ describes the response of the medium via the electric polarization $\vec{P}$ which, in our case, is generated by the phonons. We neglect any magnetic response, such that $\vec{H} = \vec{B}/\mu_0$, and therefore the energy density decomposes into two terms:
\begin{equation}
\label{eq:energydensity}
w_{\rm{EM}} = \frac{1}{2} \left(\epsilon_0 \vert \vec{E}\vert^2 + \frac{1}{\mu_0} \vert \vec{B} \vert^2 \right) +\frac{\vec{E} \cdot \vec{P}}{2}.
\end{equation}
The first and second terms of Eq.~\eqref{eq:energydensity} are the usual free-space electromagnetic energy density. For confined fields, the additional boundary conditions yield the standard quantization of the electromagnetic field in term of a discrete set of harmonic oscillator modes \cite{Dutra}. After quantization, the third term of the energy density yields a photon-phonon interaction term, with a coupling constant that depends on properties of the material hosting the phonons, and on the cavity geometry. While such a term includes contributions from all the photon and phonon modes, conservation of energy implies that only modes with matching frequencies will couple efficiently.

To describe the coupled dynamics of $\Gamma$-point phonons and cavity photons, we have then to specify the electric and magnetic field operators $\hat{\vec{E}}$ and $\hat{\vec{B}}$, which in turn depend on the cavity properties, including its geometry. We focus on the case of a simple rectangular cavity, as depicted in Fig. \ref{fig:Scheme} for which the vector potential $\vec{A}$ yields  
\begin{equation}
\vec{A} = \sum_{p,k} \sqrt{\frac{\hbar}{2 \omega_k \epsilon_0 V_c}} \left( \vec{e}_p c_{p,k} e^{i \vec{k} \cdot \vec{r}} + \vec{e}^*_p c^*_{p,k} e^{-i \vec{k} \cdot \vec{r}} \right).
\end{equation}
Each mode has an amplitude $c_{p,k}(t)= c_{p,k}(0) e^{-i \omega_k t}$ defined by a wave-vector $k$, a frequency $\omega_k$, and a polarization $p$ associated with the polarization vector $\vec{e}_p$. The factor $\sqrt{\hbar/2 \omega_k \epsilon_0 V_c}$ represents the zero-point fluctuations of the vector potential, which depend on the cavity volume $V_c$.  Assuming that the material hosting the phonons does not significantly alter the cavity modes, an approximation valid in the limit $V_s \ll V_c$, we can quantize the cavity modes following the standard procedure, i.e. by promoting the amplitudes $c_{p,k}$ to annihilation operators of bosonic modes. The electric field operator is then $\hat{\vec{E}}= - \partial_t \hat{\vec{A}}$.

The resulting phonon-cavity interaction Hamiltonian is written as follows,
\begin{equation}
\label{eq:HamCoup}
\begin{aligned}
\hat{H}_{ac} &= \int_{V_s} d^3 r \frac{\hat{\vec{E}} \cdot \hat{\vec{P}}}{2}\\
&= \sum_{\nu, p, k} i \hbar g_{k} \left[ (\vec{e}_\nu \cdot \vec{e}_{p}) e^{i k \bar{z}}  \hat{c}_{p,k} -{\rm{H.c.}} \right] \left( \hat{a}_{\nu} +\hat{a}^\dagger_{\nu} \right),
\end{aligned}
\end{equation}
where $\bar{z}$ is the position of the center of the sample. In the limit of small sample-cavity volume ratio, the phonon-cavity coupling rate $g_{\nu,k} $ becomes 
\begin{equation}
\label{eq:pccoupling}
g_{k}  = \frac{V_s Z_\nu}{4 V_0} \left[ \frac{\omega_k}{\omega_{0 \nu} \epsilon_0 V_c N_p}\right]^{1/2},
\end{equation}
which depends on $\vec{Z}_{\nu}$, the effective charge vector for the polarization, see Eq.~\eqref{eq:Pol}.
We consider that both phonon polarizations $\vec{e}_{\nu}$ have the same effective charge amplitude $Z_\nu$. Throughout the text, we will use the typical cavity and phonon parameters given in Table~\ref{Table0}, and a cavity with volume $10\,\mbox{mm}^3$. For this set of parameters, we obtain a coupling rate $g_{k} \approx 10$ GHz. While such a coupling is larger than the typical cavity decay, $\Gamma$-point phonons exhibit strong decay rates $\sim 10^{-1} \omega_{0 \nu}$, yielding a moderate phonon-photon cooperativity $\mathcal{C}  = 4g_{k}^2/\kappa_a \kappa_c \approx 10$. Such a cooperativity scales with the square of the sample volume $V_s$, and thus smaller samples will yield weaker coupling rates and lower cooperativities. Otherwise, smaller cavities yield larger coupling rates.

\begin{table}[t!]
\begin{center}
\caption{Numerical values of the parameters used in this manuscript
\label{Table0}}
\end{center}
\begin{tabular}{ |c|c|c| } 
\hline
\textbf{Parameter} & \textbf{Symbol} & \textbf{Value} \\
\hline
Phonon effective Charge & $Z_{\nu}$ & $1.54$ $\rm{e/\sqrt{a.m.u.}}$ \\
\hline
Unit cell volume & $V_0$ & $\sim 9$ ${\AA}^3$ \\
\hline
Sample volume& $V_s$ & $10 \mu {\rm{m}}^3$ \\
\hline
$\Gamma$-point phonon frequency & $\omega_{0\nu}$ &  $2 \pi \times 2.7$ THz \\
\hline
Phonon decay rate & $\kappa_a$ & $ 2 \pi \times0.27$ THz\\
\hline
Cavity frequency & $\omega_{c}$ & $2 \pi \times 2.7$ THz \\
\hline
Cavity decay rate & $\kappa_c$ & $2 \pi \times 2$ GHz \\
\hline
\end{tabular}
\end{table}

The coupling between the $\Gamma$-point phonons and a given cavity mode in Eq.~\eqref{eq:pccoupling} depends on the scalar product $\vec{e}_\nu \cdot \vec{e}_p$ between the cavity mode with polarization $p$ and the phonon mode with polarization $\nu$. Modes with orthogonal polarizations will not interact, a consequence of polarization selection rules underlying the phonon-photon interaction. Standard rectangular cavities, such as a Fabry-Perot resonator, always support two orthogonally polarized modes for each wave-vector. Nevertheless, cavities can be engineered to support only modes of a given handedness, which are called chiral cavities. In the following, we compute the phonon population that can be driven via such a chiral cavity. In the appendix, we also asses that a rectangular cavity supporting two linearly polarized modes can be used to drive chiral phonons.

\subsection{Chiral cavity modes}

Chiral cavities are designed to support modes of only one handedness. This can be achieved in a Fabry-Perot setup with mirrors that preserve circular polarization upon reflection. As a consequence, photons propagating in opposite directions are not indistinguishable, and we have to consider the contributions from positive and negative wave-vectors in the Hamiltonian \eqref{eq:HamCoup} separately. In what follows, we consider a chiral cavity that supports a mode with left-hand polarization at a frequency $\omega_c$ with a wave vector $k_c = k$. The description of a right-handed chiral cavity is analogous.

The phonon-cavity interaction Hamiltonian can be obtained from Eq.~\eqref{eq:HamCoup} (from now on we drop the subscript $k$ in the coupling rate $g$)
\begin{equation}
\label{eq:HamC0}
\begin{aligned}
\hat{H}_{\rm{Chir}} &= \sum_{\nu = x,y} i \hbar g (\vec{e}_\nu \cdot \vec{e}_{L}) \left( e^{i k \bar{z}}  \hat{c}_{L,k} + e^{-i k \bar{z}}  \hat{c}_{L,-k}  \right) \left( \hat{a}_{\nu} +\hat{a}^\dagger_{\nu} \right) \\
&\quad \quad- {\rm{H.c.}},
\end{aligned}
\end{equation}
here, the polarization vector for the left circularly polarized cavity mode is defined as $\vec{e}_L = (\vec{e}_x - i \vec{e}_y)/\sqrt{2}$. The operator $\hat{c}_{L,k}$ is associated with a mode that has a fixed handedness: it destroys a L-handed photon. As discussed above, photons propagating at opposite wave-vectors are not indistinguishable. The above Hamiltonian can be further simplified by defining the mode that couples to the phonons (the ``coupled'' mode) $\hat{c}_{\mathcal{C}} = ( e^{i k \bar{z}}  \hat{c}_{L,k} + e^{-i k \bar{z}}  \hat{c}_{L,-k})/\sqrt{2}$ and writing the phonon operators in the circularly polarized basis $\hat{a}_{L} = (\hat{a}_x + i \hat{a}_y)/ \sqrt{2}$ and $\hat{a}_{R} = (\hat{a}_x - i \hat{a}_y)/ \sqrt{2}$. Including the free cavity and phonon terms the full phonon-cavity Hamiltonian yields
\begin{equation}
\begin{aligned}
\frac{\hat{H}_{\rm{Chir}}}{\hbar} &= \omega_c\left( \hat{c}_{\mathcal{C}}^\dagger \hat{c}_{\mathcal{C}} + \hat{c}_{\mathcal{D}}^\dagger \hat{c}_{\mathcal{D}} \right) + \sum_{p = R,L} \omega_{0\nu} \hat{a}_p^\dagger \hat{a}_p  \\
&+i \hbar \sqrt{2} g \left( \hat{a}_R + \hat{a}_L^\dagger \right) \hat{c}_{\mathcal{C}} - {\rm{H.c.}} \\
& - i \left(\varepsilon e^{-i \Omega t} \hat{c}_\mathcal{C}^\dagger + \varepsilon^\prime e^{-i \Omega t} \hat{c}_\mathcal{D}^\dagger - {\rm{H.c.}} \right).
\end{aligned}
\end{equation}
Where we have defined the uncoupled mode $\hat{c}_\mathcal{D} =( e^{i k \bar{z}}  \hat{c}_{L,k}  - e^{-i k \bar{z}}  \hat{c}_{L,-k})/\sqrt{2}$, a mode that does not couple to phonons even though it has a definite (left) handedness by construction. The drive amplitudes $\varepsilon^{(\prime)}$ depend on the details of the drive scheme. We notice that only left-handed phonons exchange energy efficiently with the cavity modes: the interaction term $\hat{a}_R \hat{c}_{\mathcal{C}}$ is off-resonant. In what follows, we discard such a term with a rotating wave approximation, valid for $g \ll \omega_{0\nu,c}$, such that only phonons with the handedness matching that of the cavity will be driven.

\section{Steady-state phonon population and effective magnetic field}

The driven-dissipative dynamics of the system can be modeled with the standard Heisenberg-Langevin equation for an operator $\hat{o}$ \cite{Gardiner} 
\begin{equation}
\label{eq:HeisLang}
\partial_t \hat{o} = - \frac{i}{\hbar} [\hat{o}, \hat{H}] - \frac{\kappa_o}{2} \hat{o}  -\sqrt{\kappa_o} \hat{o}_{\rm{in}}(t),
\end{equation}
where $\hat{o}_{\rm{in}}(t)$ is an input noise operator accounting for thermal and vacuum fluctuations, with correlations $\langle \hat{o}^\dagger_{\rm{in}}(t)  \hat{o}_{\rm{in}}(t^\prime) \rangle = 1+ \langle \hat{o}_{\rm{in}}(t)  \hat{o}^\dagger_{\rm{in}}(t^\prime) \rangle = n_{{\rm{Th}},o} \delta(t - t^\prime)$ (where $n_{{\rm{Th}},o}$ is a thermal occupation given by the Bose-Einstein distribution), and $\kappa_o$ describes the total decay rate for each mode $\hat{o}$. The combination of drive and dissipation will steer the system towards a steady-state with a coherent amplitude given by $\partial_t \langle \hat{o} \rangle = 0$. The corresponding steady state number of photons  $n_{L} = \vert \langle \hat{a}_L \rangle \vert^2$ yields
\begin{equation}
\label{eq:OccLeft}
n_{L}^{\rm{Chi}} = \frac{ 2 g^2 \vert \varepsilon\vert^2}{\vert (i \Delta_\nu - \frac{\kappa_a}{2})(i \Delta_c - \frac{\kappa_c}{2}) + 2 g^2 \vert^2},
\end{equation}
where $\Delta_{\nu(c)} = \Omega - \omega_{\nu(c)}$ are the detunings between the phonon and cavity frequencies and the drive frequency. For the case considered here, the steady-state number of right-handed phonons vanishes $n_{R}^{\rm{Chi}}=0$. For a right-handed cavity, we would have a vanishing number of left-handed phonons and a number of right-handed phonons given by an expression similar to Eq.~\eqref{eq:OccLeft}. We furthermore assume that the cavity mode $\hat{c}_{\mathcal{C}}$ decays only into a single input port, such that we can write the drive amplitude $\vert \varepsilon \vert = \sqrt{\kappa_c \mathcal{P}/\hbar \Omega}$, where $\mathcal{P}$ is the laser power that drives the mode.

The driven cavity generates an imbalance between left and right-handed phonons, which in turn can be associated to an effective magnetic field $B_{\rm{eff}}$ given by \cite{shabala2024phononinverse}:
\begin{equation}
 B_{{\rm{eff}} } = \frac{\hbar \omega_{0\nu} \vert \mathcal{G} \vert^2}{g_J \mu_B (\Delta^2 -\hbar^2 \omega_{0\nu}^2)}\left((n_L - n_R)+ \frac{1}{2} \right),
\end{equation}
where $\mathcal{G}$ is the electron-phonon coupling, $\Delta$ is the gap between valence and conduction bands, $\mu_B$ is the Bohr magneton and $g_J$ is the Land\'{e} factor. For a left-handed chiral cavity, as we obtained before, the population imbalance $(n_L - n_R)  = n_{L}$ given by Eq.~\eqref{eq:OccLeft}, consequently, the effective field is proportional to the drive amplitude $\vert \varepsilon\vert^2$ which itself is proportional do the power of the laser driving the cavity. At a fixed drive power, the cavity-phonon coupling rate $g$ and the drive frequency, encoded in the detunings $\Delta_{\nu,c}$ define the optimum operation point.

We focus on the situation where the cavity is designed to exhibit a mode at resonance with the phonons, such that $ \Delta_\nu = \Delta_c $. The drive frequency that maximizes the phonon population depends on how strongly the phonons are coupled to the cavity mode. If $g<\sqrt{\kappa_a^2 +\kappa_c^2}/4$, the steady-state population of phonons is optimized for $\Delta_c = 0$ (drive at resonance). Otherwise, if $g>\sqrt{\kappa_a^2 +\kappa_c^2}/4$, the optimal drive-cavity detunings are  $\Delta_c = \pm \sqrt{2 g^2 - (\kappa_a^2 +\kappa_c^2)/8}$. In this second regime, cavity and phonons are hybridized and we expect a smaller phonon steady-sate population, as hybridization will distribute excitations between the two modes. In fact, the phonon population (and hence the effective magnetic field) is maximized for the coupling $g= g_{\rm{opt}} = \sqrt{\kappa_c \kappa_a/8}$ corresponding to a cavity-phonon cooperativity $\mathcal{C} = 1/2$. Such a cooperative is lower than the one estimated before, which could be experimentally adjusted by using a larger cavity, as the coupling scales as $V_c^{-1/2}$. In such an optimized scenario, the number of phonons generated by the driven cavity is $n_L^{\rm{max}} = \vert \varepsilon\vert^2/\kappa_a \kappa_c$, and depends on both the strength of the drive and the phonon dissipation rate.

\begin{figure}
    \centering
    \includegraphics[width=1.\columnwidth]{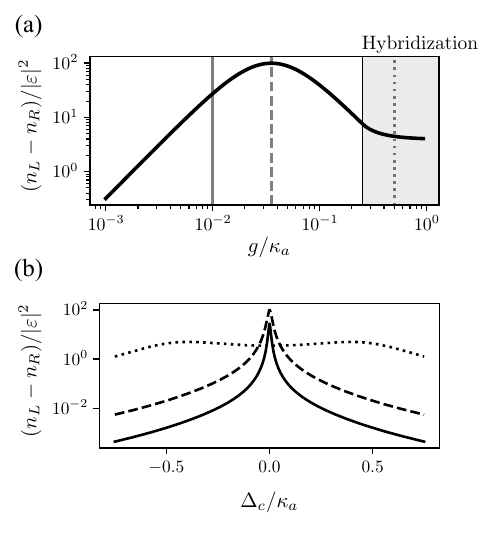}
    \caption{Imbalance between left and right-handed phonons generated by a coherently driven chiral cavity. (a) Maximum imbalance as a function of the cavity-phonon coupling. The gray area indicates $g>\sqrt{\kappa_a^2 +\kappa_c^2}/4$, the hybridized regime. (b) Imbalance as a function of the cavity-drive detuning for $g = \kappa_c$ (solid line), $g = g_{\rm{opt}} = \sqrt{\kappa_a \kappa_c/8}$ (dashed line), and $g = 3 \sqrt{\kappa_a^2 + \kappa_c^2}/8$ (dotted line). All parameters as in Table~\ref{Table0}.}
    \label{fig:Pop}
\end{figure}

In Fig.~\ref{fig:Pop}, we show the population imbalance as a function of the cavity-phonon coupling and of the cavity-drive detuning. The maximum population imbalance, indicated with a dashed line in.~\ref{fig:Pop}(a), is generated at the optimal coupling $g$. For stronger couplings, the phonons and the cavity modes hybridize, and the steady-state phonon population at a given drive amplitude is smaller. This regime is marked by the double peaked structure in Fig.~\ref{fig:Pop}(b), with maxima at detunings corresponding to the cavity-phonon polaritons frequencies.

Using the parameters of \cite{shabala2024phononinverse} and Table~\ref{Table0}, we have $B_{\rm{eff}} \approx 0.66 \left((n_L - n_R)^{l} + \frac{1}{2} \right)\,$ mT. For the optimal operation point -- the maximum of Fig.~\ref{fig:Pop} -- we can write the simplified expression for the effective field in terms of the power of the drive laser $\mathcal{P}$ as
\begin{equation}
B_{\rm{eff}} \approx (0.66 \,{\rm{mT}} ) \frac{\mathcal{P}}{ \hbar \Omega \kappa_a},
\end{equation}
which for a laser with a power $10$ nW gives an effective field $\sim 4 \, \rm{mT}$ and a coherent phonon population of $\sim 6$ phonons. Such effective field is compatible with those generated with THz pulses \cite{luo2023large,Davies2024,basini2024,Biggs2025UltrafastLiNbO3}. A key difference is that, in our proposal, the effective magnetic field is at steady-state, and not a transient effect. We also notice that our estimate is an order of magnitude smaller than the highest value measured in \cite{Biggs2025UltrafastLiNbO3}.

\subsection{Measurement considerations}

Once the phonons are generated, their fingerprints could be measured in different ways. Monitoring the cavity transmission (or reflection), can provide information about the cavity-phonon coupling but, since the system is linear, would not be a direct probe of the amount of phonons generated. For this, the best is to use an auxiliary laser, with a frequency detuned from the phonon frequency. Brillouin light scattering from the chiral phonons can then be used to infer their population, as it is done in experiments measuring the Faraday rotation angle generated by phonons \cite{basini2024}. Alternatively, the material hosting the phonons can be connected to cantilevers, such that the torque generated by the chiral phonons can be directly probed as was done in \cite{Zhang_2025_Measurement}. The drawback of this approach, is the additional phonon dissipation due to the clamping to the cantilevers.

Another important aspect to be considered is the thermal and vacuum noise introduced by the cavity which might change the incoherent phonon population. In this sense, the phonon number fluctuations will be modified, which in turn will imply modified fluctuations of the effective magnetic field generated by the chiral phonons. Such an effect has different traits depending on whether the cavity is linear or chiral, and could be measured via the number fluctuations of the cavity output, which can be accomplished via intensity noise measurements, or by second-order coherence function of the output field in a Hanbury Brown–Twiss setup \cite{Scully1997QuantumOptics}. Both techniques have been staples for probing phonon with photons from a cavity \cite{Purdy2013RadiationPressureShotNoise,Cohen2015PhononCounting}, although in a different coupling regime.

\section*{Conclusions}

Generating chiral phonons often rely on short pulsed lasers that drive a transient population of phonons in a sample. Here, we proposed an alternative method, in which confinement of the electromagnetic field, via a cavity, can be used to generate a steady-state population of chiral phonons. Our idealized setup consists of a cavity supporting a mode with only one handedness, a so-called chiral cavity, such that only phonons with the matching handedness will couple to the cavity mode. By coherently driving the cavity mode (via an external port), a steady-state population of chiral phonons is generated. 

For state-of-the-art parameters of cavities, and for illustrative material parameters, we showed that a coherent steady-state population of~$6$ phonons can be generated with a drive laser with 10 nW of power. This corresponds to an effective phonon-generated magnetic field of $~4$ mT, in line with the effective fields obtained in experiments with pulsed THz light \cite{luo2023large,Davies2024,basini2024,Biggs2025UltrafastLiNbO3}, albeit being an steady-state field. In the appendix we also show an alternative setup that uses a linear cavity with two degenerate modes that are linearly polarized. In this case, both modes have to be driven simultaneously with drive amplitudes that are $\pi/2$ out of phase, and achieving maximum phonon occupancy requires a larger phonon-photon cooperativity. Cavities that exhibit modes carrying orbital angular momentum, e.g. Laguerre-Gauss modes, could also be used to generate chiral phonons, a framework that we postpone for a future work.

\section*{Acknowledgements}

VASVB thanks Quansheng Zhang for useful discussions.

NS and RMG acknowledge support from the Swedish Research Council (VR starting Grant No. 2022-03350), the Olle Engkvist Foundation (Grant No. 229-0443), the Royal Physiographic Society in Lund (Horisont), the Knut and Alice Wallenberg Foundation (Grant No. 2023.0087), and Chalmers University of Technology, via the department of physics and the Areas of Advance Nano and Materials Science. 

\section*{Appendix}

\subsection*{Linear cavity driving chiral phonons}

\begin{figure}[h]
    \centering
    \includegraphics[width=0.75\columnwidth]{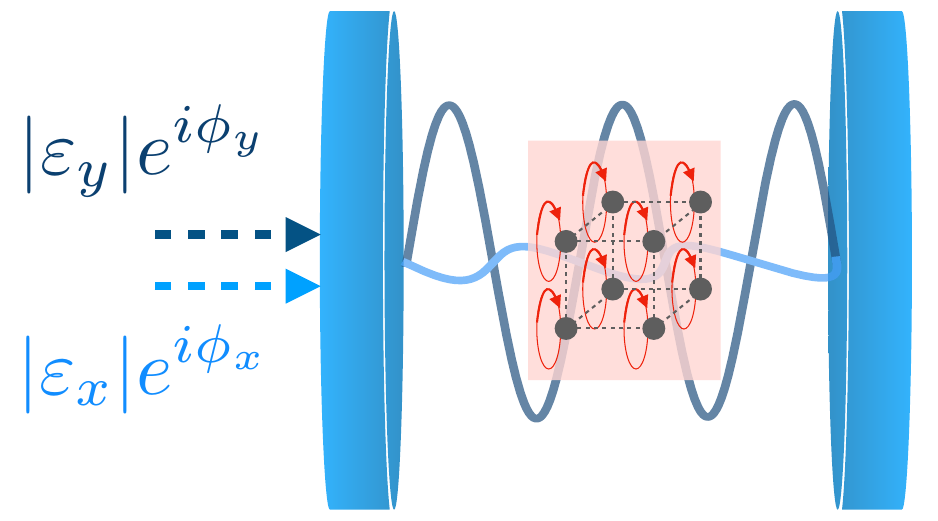}
    \caption{Chiral Phonons generated by a coherently driven linear cavity. In this case, the cavity has to support two degenerate modes that are linearly polarized. Such modes have to be driven with independent drives that have a relative phase.}
    \label{fig:SchemeLin}
\end{figure}

While in the main text we focused on a chiral cavity, i.e. a cavity supporting perfectly circularly polarized modes, cavities supporting linearly polarized modes can also be used to generate a steady-state population of chiral phonons. In this case, the cavity is required to exhibit two degenerate and linearly polarized modes that are driven with independent external drives, as depicted in Fig.~\ref{fig:SchemeLin}.

\vspace{0.5cm}

Let's consider the hypothetical case of a linear cavity supporting two degenerate modes with polarization $\vec{e}_p = \vec{e}_{x,y}$. In such cavities, we can consider the contribution from only one wave-vector for each mode, and thus the Hamiltonian in Eq.~\eqref{eq:HamCoup} simplifies to
\begin{equation}
\label{eq:HamL0}
\hat{H}_{L} = i \hbar g \sum_{p = x, y} ( e^{i k \bar{z}}\hat{c}_p - e^{-i k \bar{z}} \hat{c}^\dagger_p) (\hat{a}_p +\hat{a}^\dagger_p).
\end{equation}
Notice that the coupling rate $g$ is independent of the direction of polarization of the cavity field.
We further notice that, in contrast with the chiral cavity case, the phase factor $e^{i k \bar{z}}$ is also independent of the polarization $p$ and can be eliminated via a gauge transformation $\hat{c}_p \rightarrow \hat{c}_p e^{- i k \bar{z}}$, or, alternatively, by fixing the sample position such that $k \bar{z} = 2 m \pi$ (with $m$ and integer). The Hamiltonian in Eq.~\eqref{eq:HamL0} can be written in terms of right and left-handed modes $\hat{o}_{R,L} = (\hat{o}_x \mp i \hat{o}_y)/\sqrt{2}$ ($o = c,a$) as
\begin{equation}
\label{eq:HamL1}
\begin{aligned}
\hat{H}_{L,I} &= i \hbar g \left[ (\hat{c}_R - \hat{c}_L^\dagger)(\hat{a}_R^\dagger + \hat{a}_L) - {\rm{H.c.}} \right].\\
\end{aligned}
\end{equation}
The cavity-phonon Hamiltonian is therefore given by
\begin{equation}
\begin{aligned}
\frac{\hat{H}_{\rm{Lin} }}{\hbar} &=  \sum_{p = R,L}\left[\omega_c \hat{c}_{p}^\dagger \hat{c}_{p} + \omega_{0\nu} \hat{a}_p^\dagger \hat{a}_p  \right] \\
&+ i \hbar g \left[ (\hat{c}_R - \hat{c}_L^\dagger)(\hat{a}_R^\dagger + \hat{a}_L) - {\rm{H.c.}} \right] \\
& - i \left(\varepsilon_R e^{-i \Omega t} \hat{c}_R^\dagger + \varepsilon_L e^{-i \Omega t} \hat{c}_L^\dagger - {\rm{H.c.}} \right).
\end{aligned}
\end{equation}
The last term describes external coherent drives with a frequency $\Omega$ that are applied to both $x$ and $y$ polarized mode: $\varepsilon_{R,L} = (\varepsilon_x \pm i \varepsilon_y)/\sqrt{2}$, where $\varepsilon_p = \vert \varepsilon_{x,y} \vert e^{i \phi_{x,y}}$. The driven-dissipative dynamics of the system can be modeled by standard Heisenberg-Langevin equation, as done in the main text. In this case, the coherent steady-state phonon population $n_{L(R)} = \vert \langle \hat{a}_{L,R}\rangle \vert^2$ is
\begin{equation}
n_{L(R)}^{\rm{Lin}} = \frac{g^2 [\vert \varepsilon_x\vert^2 + \vert \varepsilon_y\vert^2 \pm 2 \vert \varepsilon_x\vert \vert \varepsilon_y\vert \sin(\phi_x - \phi_y) ] }{2\vert (i \Delta_\nu - \frac{\kappa_a}{2})(i \Delta_c - \frac{\kappa_c}{2}) + g^2 \vert^2},
\end{equation}

The population difference $(n_L - n_R)^{\rm{Lin}}$ is now given by
\begin{equation}
\label{eq:ImbLin}
\left(n_{L}-n_{R} \right)^{\rm{Lin}} = \frac{g^2 \vert \varepsilon_{x} \vert \vert \varepsilon_{y} \vert {\rm{sin}} \left( \phi_x - \phi_y \right)}{\vert (i \Delta_\nu - \frac{\kappa_a}{2})(i \Delta_c - \frac{\kappa_c}{2}) +g^2 \vert^2},
\end{equation}
and hence a net effective magnetic field, or phonon angular momentum, requires that $\phi_x - \phi_y \neq N \pi$. Specifically, $B_{\rm{eff,Lin}}$ is maximized for $\phi_x - \phi_y = \pm (2N+1)\pi/2$ (out-of-phase drives). In the case where both linear polarizations are driven with the same amplitude $\vert \varepsilon_x \vert = \vert \varepsilon_y \vert$, such an effective field is entirely due either left and right-handed phonons depending on whether $\phi_x - \phi_y =\pi/2$ or $-\pi/2$ respectively. We also notice that the phonon population imbalance is maximized, in this case, for $g = \sqrt{\kappa_a \kappa_c}/2$, and a phonon-cavity cooperativity of $\mathcal{C} = 1$, and thus requires a stronger coupling than the chiral cavity presented in the main text.

\bibliography{references} 
\end{document}